\documentclass[a4paper,fleqn,usenatbib]{mnras}
\usepackage{graphicx}	
\usepackage{amsmath}	
\usepackage{amssymb}	
\usepackage{deluxetable}
\usepackage{multicol}

\def\deg {$^{\circ}$}

\newcommand{\HII}{H$\,${\sc ii}}

\def\kms  {km\,s$^{-1}$}
\def\mo   {\ifmmode{{\rm M}_{\odot}}\else{M$_{\odot}$}\fi}

\def\bco {\ifmmode{^{12}{\rm CO}(J=2\to1)}\else{$^{12}{\rm
CO}(J=2\to1)$}\fi}
\def\m  {\ifmmode{\mu {\rm m}}\else{$\mu$m}\fi}
\def\cco {\ifmmode{^{13}{\rm CO}(J=1\to0)}\else{$^{13}{\rm
CO}(J=1\to0)$}\fi}
\def\dco {\ifmmode{^{13}{\rm CO}(J=2\to1)}\else{$^{13}{\rm
CO}(J=2\to1)$}\fi}
\def\eco {\ifmmode{{\rm C}^{18}{\rm O}(J=1\to0)}\else{{\rm C}$^{18}{\rm
O}(J=1\to0)$}\fi}
\def\hi  {{H\ts {\scriptsize I}}\fi}

\def\Hb  {\ifmmode{{\rm H}{\alpha}}\else{H\ts {$\beta$}}\fi}

\def\nh  {\ifmmode{N(\hi)}\else{$N$(\hi)}\fi}
\def\hun  {\ifmmode{I_{100}}\else{$I_{100}$}\fi}
\def\sex  {\ifmmode{I_{60}}\else{$I_{60}$}\fi}
\def\hh   {\ifmmode{{\rm H}_2}\else{H$_2$}\fi}
\def\nhh   {\ifmmode{N({\rm H}_2)}\else{$N$(H$_2$)}\fi}
\def\zwco  {\ifmmode{^{12}{\rm CO}}\else{$^{12}{\rm CO}$}\fi}
\def\nzwco  {\ifmmode{N(^{12}{\rm CO})}\else{$N(^{12}{\rm CO})$}\fi}
\def\wzwco  {\ifmmode{W(^{12}{\rm CO})}\else{$W(^{12}{\rm CO})$}\fi}
\def\drco  {\ifmmode{^{13}{\rm CO}}\else{$^{13}{\rm CO}$}\fi}
\def\ndrco  {\ifmmode{N(^{13}{\rm CO})}\else{$N(^{13}{\rm CO})$}\fi}
\def\wdrco  {\ifmmode{W(^{13}{\rm CO})}\else{$W(^{13}{\rm CO})$}\fi}
\def\tex  {\ifmmode{T_{ex}({\rm CO})}\else{$T_{ex}({\rm CO})$}\fi}
\def\xco   {\ifmmode{X_{\rm CO}}\else{$X_{\rm CO}$}\fi}
\def\msol   {\ifmmode{{\rm M}_{\odot}}\else{M$_{\odot}$}\fi}
\def\amm    {NH$_{3}$}

\def\ha     {H$\alpha$}

\def\methanol {CH$_3$OH}
\def\water {H$_2$O} 
\def\sfr {M$_{\odot}$\,yr$^{-1}$}
\def\Lsun{L$_{\odot}$}
\def\Msun{M$_{\odot}$}

\title[Nuclear Starburst Diagnostics:\water\ and \methanol]{Diagnostics of a Nuclear Starburst: Water and Methanol Masers}

\author[M. D. Gorski et al.]{
Mark D. Gorski,$^{1}$\thanks{E-mail: mgorski3@uwo.ca},
J\"urgen Ott,$^{2}$\thanks{E-mail: jott@nrao.edu},
Richard Rand$^{3}$\thanks{E-mail: rjr@unm.edu},
David S. Meier$^{4,2}$,
\newauthor{Emmanuel Momjian$^{2}$,
Eva Schinnerer$^{5}$, \&
Simon P. Ellingsen$^{6}$}
\\
$^{1}$Department of Physics and Astronomy, University of Western Ontario, 1151 Richmond Street, London, Ontario, N6A 3K7, Canada\\
$^{2}$National Radio Astronomy Observatory, P.O. Box O, 1003 Lopezville Road, Socorro, New Mexico, 87801, USA\\
$^{3}$Department of Physics and Astronomy, University of New Mexico, 1919 Lomas Boulevard NE, Albuquerque, New Mexico, 87131, USA\\
$^{4}$Department of Physics, New Mexico Institute of Mining and Technology, 801 Leroy Place, Socorro, New Mexico, 87801, USA\\
$^{5}$ Max-Planck Institut f\"ur Astronomie, K\"onigstuhl 17, D-69117 Heidelberg, Germany\\
$^{6}$ School of Natural Sciences, University of Tasmania, Hobart TAS 7001, Australia
}

\date{Accepted XXX. Received YYY; in original form ZZZ}

\pubyear{2018}

\begin{document}
\label{firstpage}
\pagerange{\pageref{firstpage}--\pageref{lastpage}}
\maketitle

\begin{abstract}

We test models of starburst driven outflows using observations of the 22.2\,GHz \water\  and 36.2\,GHz class I \methanol\ maser lines. We have observed the starburst galaxy NGC\,253 using the Karl G. Jansky Very Large Array.  We present evidence for entrainment of star-forming dense-molecular gas in the outflow of NGC\,253. We also show that \water\ masers are associated  with forming super star clusters and not with supernova remnants.  We detect four new 36\,GHz \methanol\ masers in the central kpc and show possible evidence for a star-formation origin of two class I \methanol\ masers. Such high resolution observations are essential for understanding the origin of these masers.   
 
\end{abstract}


\section{Introduction}

 Feedback from stars is critical to galaxy evolution in order to slow down star formation and remove baryons from the galactic disk (e.g., \citealp{Hopkins2012}). Galactic winds are one particular mechanism thought to slow the formation of stars over cosmic time. They potentially do this by removing large amounts of  molecular gas from the disk of star-forming galaxies, i.e., ejective feedback (e.g., \citealp{Somerville2015}). Starburst driven winds are a result of feedback in the form of supernovae and stellar winds (e.g., \citealp{Veilleux2005, Li2017}).  In this paper we  explore the nuclear starburst of the nearby galaxy NGC\,253 using astrophysical \water\ and \methanol\ masers.

NGC\,253 lies at a distance of 3.5\,Mpc measured from the planetary nebulae luminosity function \citep{Rekola2005}. It has a total star-formation rate of 4.2\,\sfr\ of which $\sim50\%$ is concentrated in the nuclear starburst \citep{Leroy2015}. The nuclear region is $\sim$1\,kpc in diameter and contains about 90\% of the CO luminosity of the galaxy \citep{Young1995}. The intense star formation at the center drives a molecular outflow of $\sim$30\,\sfr\ \citep{Bolatto2013}. 

The molecular gas in the central kpc is also highly turbulent as shown by large line widths of molecular clouds in the central kpc \citep{Leroy2015}. The clouds are located in a $\sim$1\,kpc long molecular bar. The central kpc also hosts a rich spectrum of molecular lines at mm and cm wavelengths (e.g., \citealp{Martin2006,Meier2015,Gorski2017}). These molecular lines are useful tracers of physical conditions in the molecular interstellar medium (ISM). Utilizing the ALMA 3mm band, \citet{Meier2015} detect 50 different spectral lines towards the central kpc of NGC\,253 that are used to identify different conditions from photon dominated regions (PDRs), strong and weak shocks, and gas densities and opacities.
NGC\,253 is the prototypical starburst used for high redshift studies. The proximity of NGC\,253 makes detailed studies of its galactic wind and nuclear starburst possible. \citet{Westmoquette2011} model the approaching (southern) side of the bi-conical \ha\ emitting outflow with a frustum (truncated cone) aligned along the minor axis of the galaxy and with an opening angle of 60\deg. They determine that at its greatest height above the disk the \ha{} cone has a diameter of $\sim1$\,kpc and a recessional velocity of a few 100\,\kms\ relative to the systemic velocity for NGC\,253 (234 \kms \citealp{Whiting1999}).  \citet{Strickland2000,Strickland2002} studied the X-ray and \ha{} emission from the nuclear starburst of NGC\,253. From their analysis, they propose four models that can give rise to the outflow morphology seen in X-ray and \ha{} emission (see figure 11 in \citealp{Strickland2002}). They predict various degrees of entrainment of material from the disk. Model (a) shows an interaction of hot outflowing gas with cool ambient halo clouds. These cool clouds come from accretion of primordial material or  gas ejected as  galactic fountains \citep{Norman1989}. Models (b) and (c) imply entrainment of dense gas from the disk with differences in the properties of the X-ray emitting gas. (b) shows sheaths of X-ray emitting material around dense cloud cores, whereas (c) shows a shell of X-ray emitting gas encompassing the entire outflow. In the last model, (d), the hot outflow interacts with the galactic disk and generates a shell of cooler shocked material around the hot ionized outflow. If there is entrained star-forming gas it may be traced by maser emission as tentatively shown by \water\ masers extended along the axis of the outflow as reported by \citet{Gorski2017}. This would support models (b) and (c) of \citet{Strickland2002} indicating entrained gas from the disk. Sites where the hot ionized outflow is collimated out of the disk may be traced by \methanol\ masers \citep{Ellingsen2014} supporting model (d) of \citet{Strickland2002}.

In this paper we focus on \water\ and \methanol\ masers as diagnostic tools for studying the nuclear starburst and outflow in NGC\,253. \citet{Gorski2017} detect 17 spectral lines of which they focus on three species: \water\,$6_{16}-5_{23}$ (22.2\,GHz), the metastable \amm(J=K) lines (23.6945\,GHz$-$27.4779\,GHz), and \methanol\,$4_1-3_0\,E$ (36.2\,GHz). \water\,$6_{16}-5_{23}$ and \methanol\,$4_1-3_0\,E$\ are well known maser lines. Masers provide a unique opportunity to probe star-forming environments. \citet{Gorski2017} showed evidence for \water\ maser emission along the axis of the hot ionized outflow. They also resolved the first extra galactic 36\,GHz \methanol\ masers from \citet{Ellingsen2014} into five, unusually luminous sources at the edges of the molecular bar. 

The 22\,GHz \water\ line requires dense gas $>10^7$\,cm$^{-3}$ and kinetic temperatures $>\,300$\,K to mase if it is collisionally pumped. The line can also be radiatively pumped with background infrared radiation with temperatures of $\sim1000$\,K \citep{Gray2016}.  22\,GHz water masers are classified into three groups: the stellar class ($L<0.1$\,\Lsun), kilomasers ($0.1$\,\Lsun\,$<L<1$\,\Lsun{}), and megamasers (L$>20$\,\Lsun) (see \citealt{Hagiwara2001} for a description of this nomenclature). The stellar and kilomaser classes are mostly associated with star formation (e.g., \citealt{Hagiwara2001, Walsh2011, Tarchi2012}). \citet{Walsh2011} estimate that  90\% of these stellar class water masers are associated with high mass star formation (Young Stellar Objects, YSOs) and the other 10\% associated with either low mass star formation or evolved stars.  Their survey is sensitivity limited with 50\% completeness at a flux limit of 5.5\,Jy.  We will use the 22 GHz \water\ maser as a sign of intense star formation and hot shocked gas in our analysis of NGC\,253's nuclear starburst and outflow. 

\methanol\ masers are divided into two classes defined by their pumping schemes. Class I is collisionally pumped, such as the 36\,GHz transition, and class II is radiatively pumped \citep{Menten1991}.  These  masers trace shocks and dense gas $>10^4$\,cm$^{-3}$ \citep{Pratap2008}. The first detection of 36\,GHz \methanol\ masers in the extra-galactic context was by \citet{Ellingsen2014} in NGC\,253. To date there have been detections in five galaxies other than the Milky Way: NGC\,253 \citep{Ellingsen2014}, Arp220 \citep{Chen2015}, NGC\,4945 \citep{McCarthy2017}, IC\,342, and NGC\,6946 \citep{Gorski2018}. The morphology of the 36\,GHz \methanol\ emission in NGC\,253 and IC\,342 is similar to that of the HNCO molecule a weak shock tracer \citep{Gorski2017,Gorski2018}. \citet{Ellingsen2017} and \citet{McCarthy2017} interpret the emission as coming from large scale flows resulting in cloud-cloud collisions in NGC\,253 and NGC\,4945. Here we use the 36\,GHz \methanol\ maser as a tracer of shocks where the molecular outflow meets the disk in NGC\,253.

In \S\,\ref{section obs} we  describe our observations and methods for detecting masers. \S\,\ref{section res} we  report the measurements of \water\ and \methanol\ masers in the nuclear region and outflow of NGC\,253. In \S\,\ref{section disc} we  discuss the detected maser luminosities, positions, and velocities as relevant to models by \citet{Strickland2002} and forming super star clusters from \citet{Leroy2018}. Lastly, in \S\,\ref{section sum} we summarize our findings.

\section{Methods and Observations}
\label{section obs}

\subsection{Observations}
We have acquired Karl G. Jansky Very Large Array (VLA)\footnote{The National Radio Astronomy Observatory is a facility of the National Science Foundation operated under cooperative agreement by Associated Universities, Inc.} observations of the nuclear starburst of NGC\,253 (Project code: 16B-337). The VLA was in A configuration. We targeted the  22.2\,GHz \water\ and 36.2\,GHz \methanol\ maser lines.  For the 22\,GHz observations we acquired 2\,hours 40\,minutes of integration time on source and  1\,hours 30\,minutes for the 36\,GHz observations. To observe the \water\ line we centred  a 64\,MHz wide sub-band at 22.216 GHz with 4096 channels (15.6\,kHz  or 0.21\,\kms per channel). For the \methanol\ line we centred a 128\,MHz wide sub-band at 36.130\,GHz with 6144 channels ( 20.8\,kHz or 0.17\,\kms per channel). We also acquired 1\,GHz of continuum bandwidth centred at 33.5\,GHz with 1.0\,MHz (8.9\,\kms) channel widths. The data were calibrated in CASA 4.7.1 \citep{McMullin2007} using the CASA scripted pipeline version 1.3.9\footnote{https://science.nrao.edu/facilities/vla/data-processing/pipeline/scripted-pipeline}. We observed 3C48 as the flux density calibrator, J2253+1608 as the bandpass calibrator, and J0120-2701 as the complex gain calibrator. We then self-calibrated the data in phase only using a continuum image made from line-free channels .  All the image cubes were CLEANed to 3$\sigma$ rms noise with natural weighting. All velocities  are in the LSRK frame unless stated otherwise. All positions are relative to the phase tracking center (J0210-2701) with an uncertainty of 0.002\arcsec\ RA: 01$^h$20$^m$31.663334$^s$ Dec: -27\deg01\arcmin24.652570\arcsec\  . The final rms of the 33.5\,GHz continuum image is 0.02 mJy\,beam$^{-1}$. The rms of the image cubes is discussed in the \S \ref{maser id}. 

For \water, we have imaged the entire primary beam out to the half power width ($\sim$1.5\arcmin\ in diameter). We did this to capture potential masers at large heights above the disk of NGC\,253. 
Due to the large data sizes we could not image the entire primary beam in one image cube.  
We made 25 image cubes with dimensions of 800\,pix$\times$800\,pix$\times$1640\,channels. 
We smoothed the image cubes to a common spatial resolution of 0.23\arcsec$\times$0.13\arcsec\ with a position angle 20\deg\ east of north. The image cubes were made with 0.03\arcsec\ pixels.  
We binned the spectral axis to a spectral resolution of 0.5\,\kms. 

For \methanol, we made three image cubes with 0.015\arcsec\ pixels. We did not search the entire primary beam, because in the $\sim$100\,pc resolution observations from \citet{Gorski2017} there are no observed masers with anomalous velocities or positions relative to the molecular bar (\citealp{Ellingsen2014} and \citealp{Gorski2017}). Each image cube is 1200\,pix$\times$1200\,pix$\times$500\,channels. Their centers are listed in Table \ref{tab:methpos}. 
\begin{table}
\centering
\caption{ \methanol\ Cube Centers}
\label{tab:methpos}
    \begin{tabular}{ll}
    \hline
    RA (J2000)	& DEC(J2000) \\
    \hline
    00$^h$47$^m$33.890$^s$  & -25$^\circ$17\arcmin13.350\arcsec \\
    00$^h$47$^m$33.043$^s$  & -25$^\circ$17\arcmin18.357\arcsec \\
    00$^h$47$^m$32.090$^s$  & -25$^\circ$17\arcmin25.573\arcsec \\
    \hline
    \end{tabular}
\end{table}
 These locations are chosen to cover the five \methanol\ masers from \citet{Gorski2017} and the central molecular zone. We smoothed the image cubes to a common beam size with an angular resolution of 0.13\arcsec$\times$0.10\arcsec\ and a position angle of 6\deg\ east of north, and binned the velocity axis to a resolution of 1\,\kms.  We also generated a continuum image of the central kpc centered at 33.5\,GHz with bandwidth of 1\,GHz, a native resolution of 0.096\arcsec$\times$0.045\arcsec, and a position angle of 2.1\deg.

\subsection{Maser Identification}
\label{maser id}

Finding masers over large areas of sky is a question of how many real sources can be reliably identified without claiming spurious sources as real ones. \citet{Walsh2012} have provided a method for detecting galactic masers in the \water\ Southern Galactic Plane Survey (HOPS). They use a combination of visual inspection and the source finding algorithm DUCHAMP\footnote{https://www.atnf.csiro.au/people/Matthew.Whiting/Duchamp/} \citep{Whiting2012}. We use a slightly different version of \citet{Walsh2012}'s detection method, modified for our data set. 

\begin{figure}
    \centering
    \includegraphics[width=0.5\textwidth]{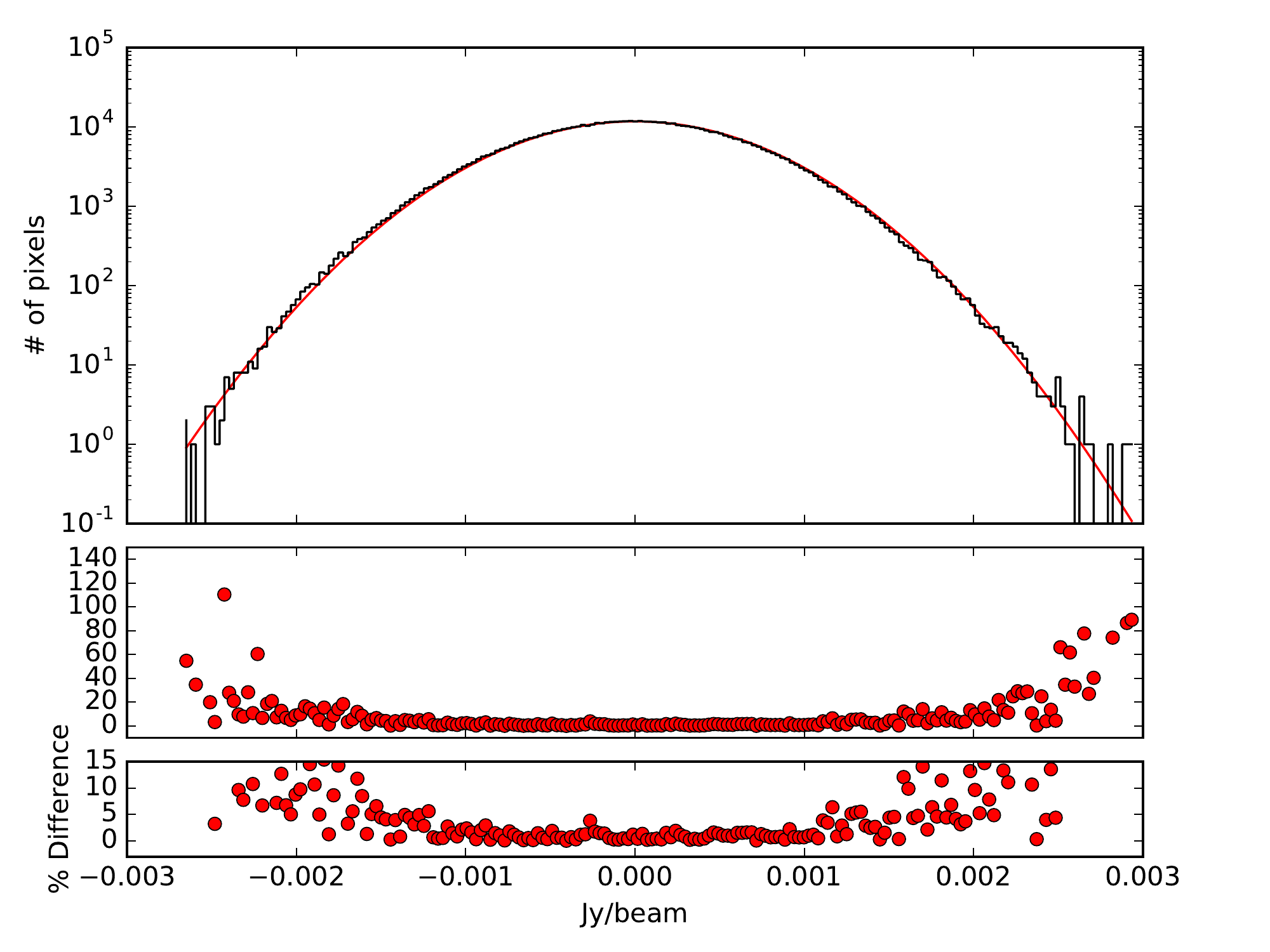}
    \caption{The top panel shows the distribution of pixel values in a maser free channel in a 22\,GHz image cube using 200 bins. The red line shows the Gaussian fit to the negative half of the data. The fit is extended to the positive half. The peak is centred at 1.2$\times10^{-6}$\,Jy and $\sigma=1.3$\,mJy. The bottom panel shows the percent difference of the data and the Gaussian fit.}
    \label{fig:dist}
\end{figure}

As narrow masers ($<0.5$\,\kms) can appear almost identical to spurious noise we first measure the rms of the image cube from the negative fluxes. This provides us with a meaningful way to quantify the quality of the data without accidentally including real sources or misidentifying real sources as noise. From the image cube we extracted a histogram of pixel values ignoring 10\% of channels at the band. We fit a Gaussian to the  negative half of the distribution of pixel values in each channel. Figure \ref{fig:dist} shows the fit extended to the positive half of the distribution for a single channel. The standard deviation in the \water\ cubes is measured from the FWHM and has an average value of 1.3$\pm$0.1\,mJy\,beam$^{-1}$ across all channels. The uncertainty is measured from the minimum and maximum values from all channels. Notably the absolute minimum value from the negative pixels over all channels is -8.8\,mJy\,beam$^{-1}$ or -6.5$\sigma$. This will be relevant for detection criteria later. For the \methanol\ data  cubes the standard deviation is 1.5$\pm$0.1\,mJy\,beam$^{-1}$.

Masers in the Milky Way are mostly point sources \citep{Walsh2014}. At the distance of NGC\,253 we therefore do not expect any spatially resolved sources given our resolution of $\sim4$\,pc. This means we assume all sources in our data should appear beam shaped. We now consider two cases. The first case is where the maser is spectrally resolved or partially spectrally resolved. The second case is where the maser is completely spectrally unresolved resulting in a single bright channel. 

In the spectrally resolved or partially-spectrally resolved case we impose the condition that the maser must appear in three or more channels and have a peak flux density of at least 5$\sigma$ with two 3$\sigma$ adjacent channels. The number of pixels in the area sustained by the half power of the synthesized beam is $\sim38$ pixels. A 5$\sigma$ point source convolved with a Gaussian 38\,pixel beam will fill $\sim27$ pixels above 3$\sigma$. We therefore impose the condition that the source must fill at least 27 pixels in the brightest channel above 3$\sigma$ and at least 45 voxels (a voxel is 1 spatial pixel by 1 spectral pixel). The condition of 45 voxels is chosen because we expect a 3$\sigma$ source to fill $\sim9$ pixels above 2.5$\sigma$, and we expect $\sim1$ source over all 25 image cubes with a 5$\sigma$ bright channel adjacent to two 3$\sigma$ channels.  This setup provided us with 10's of sources per image cube that could be visually inspected.

In the totally spectrally unresolved case we have two subsets of criteria. Either may be satisfied to be considered a spectrally unresolved source.  First, because there are no negative fluxes below -6.5$\sigma$, we consider any positive pixel with a value above 6.5$\sigma$ as a candidate source. The source must also consist of at least 38 pixels (the beam area) with values greater than 3$\sigma$. This is because a 6.5$\sigma$ source convolved with a Gaussian beam fills at least 38\,pixels above 3$\sigma$ and a spurious bright pixel will not.  This is slightly different from \citet{Walsh2012} where they chose a peak flux density value of 8$\sigma$. We do not impose any condition on the position angle (PA).  Second we choose sources with peak flux densities $>6\sigma$ and consisting of 38 pixels $>3\sigma$, We chose a $>6\sigma$ limit as we expect $\sim1$ source in the entire primary beam and entire spectral bandwidth should the noise be well described by a Gaussian. We do this by counting the number of beams, not pixels, in the image as interferometer noise is correlated \citep{Greisen2002}. We include the criterion that a detected source must have a Gaussian fit with the same PA as the synthesized beam. From our results in \S\ref{section res} we determine the range of acceptable PAs to be 20.0$\pm$7.8\deg. The uncertainty in position angle is determined from the maximum deflection from 20\deg of the four sources we detect with peak fluxes $>6.5\sigma$. This is a new constraint we impose, compared to \citet{Walsh2012}. 

The program DUCHAMP was run twice on the same image cube: once looking for spectrally resolved or partially resolved sources, and a second time looking for spectrally unresolved sources using the constraints defined above. Each list of sources was then visually inspected to eliminate any spurious detections (e.g. groups of pixels with irregular borders; figures 2 and 3 from \citealp{Walsh2012}). What remains we consider a detection of a maser.

\section{Results}
\label{section res}

\subsection{\water\ Masers}

\begin{figure*}
    \centering
    \includegraphics[width=0.495\textwidth]{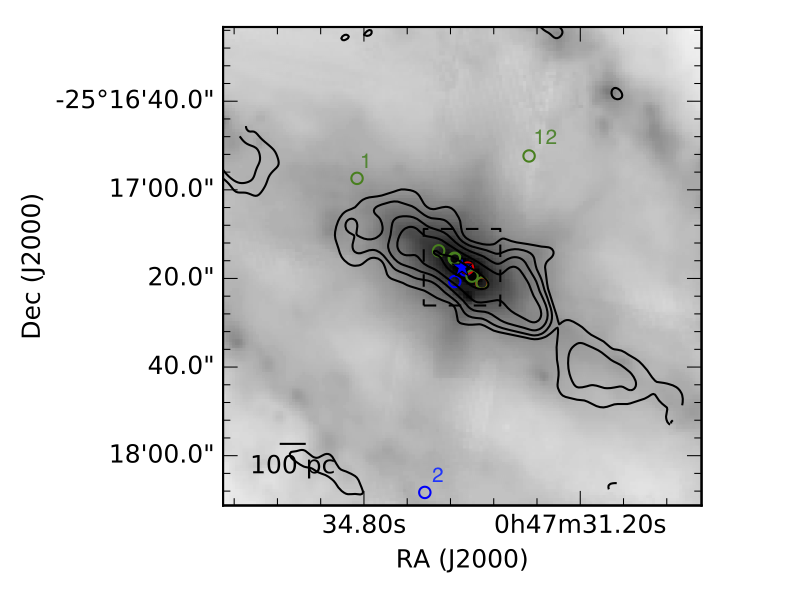}
    \includegraphics[width=0.495\textwidth]{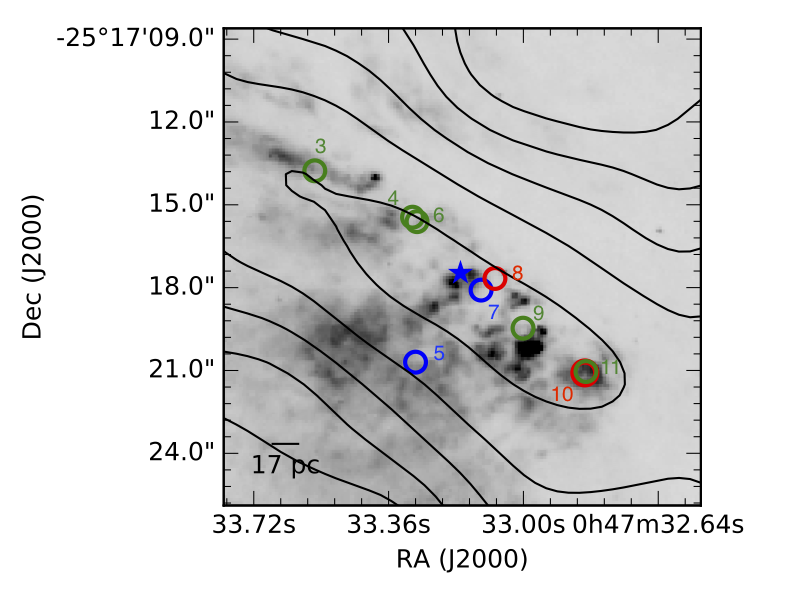}
    \caption{Positions of the 13 \water\ masers we have detected in NGC\,253. In both panels we show $\sim$4\arcsec\ resolution 60, 120, 240, and 480 Jy\,beam$^{-1}$ \kms\ contours of $^{12}$CO\,($1-0$) from \citet{Bolatto2013}. The bright kilomaser W1 is indicated by the star. Masers with velocities consistent with the molecular bar (170$-$300\,\kms) are plotted in green. Masers redshifted and blueshifted with respect to the molecular bar are respectively plotted in red and blue. Left: Shows IRAC 8$\mu$m image \citep{Dale2009} of the central $\sim1500$\,pc of NGC\,253. The box shows the region in the right panel. Right:  The nuclear 300\,pc with the HST WFPC2 H$\alpha$ map from \citet{Watson1996}.}
    \label{fig:loc}
\end{figure*}

\begin{figure*}
    \centering
    \includegraphics[width=0.95\textwidth]{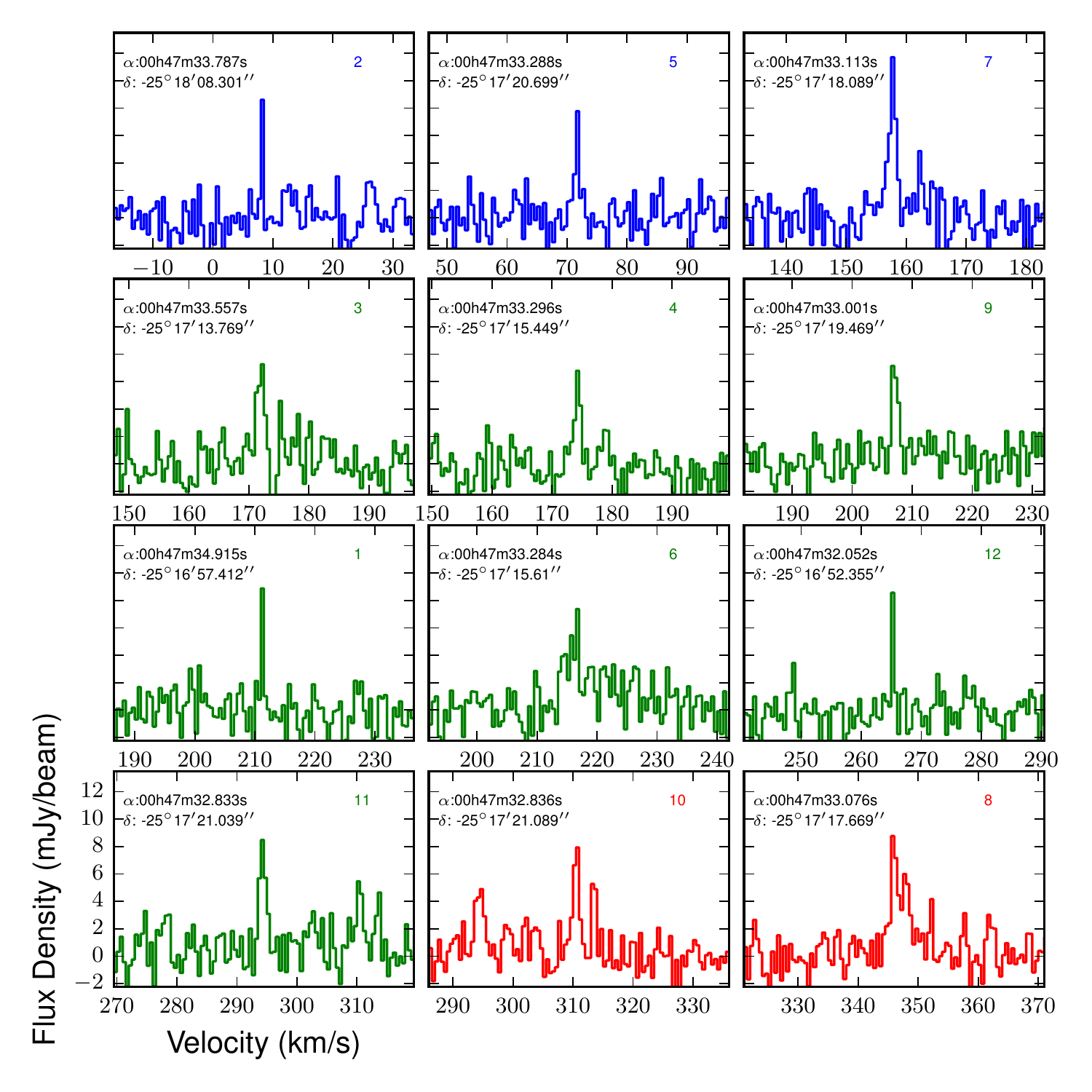}
    \caption{Spectra of the twelve stellar class \water\ masers detected besides W1 (the nuclear kilomaser). The spectral resolution is 0.5\,\kms. The coordinates of each maser is shown in the upper left corner of each panel. The spectra are colour coded as in Figure \ref{fig:loc}. These spectra are not primary beam corrected. }
    \label{fig:waterspec}
\end{figure*}

\begin{figure}
    \centering
    \includegraphics[width=0.5\textwidth]{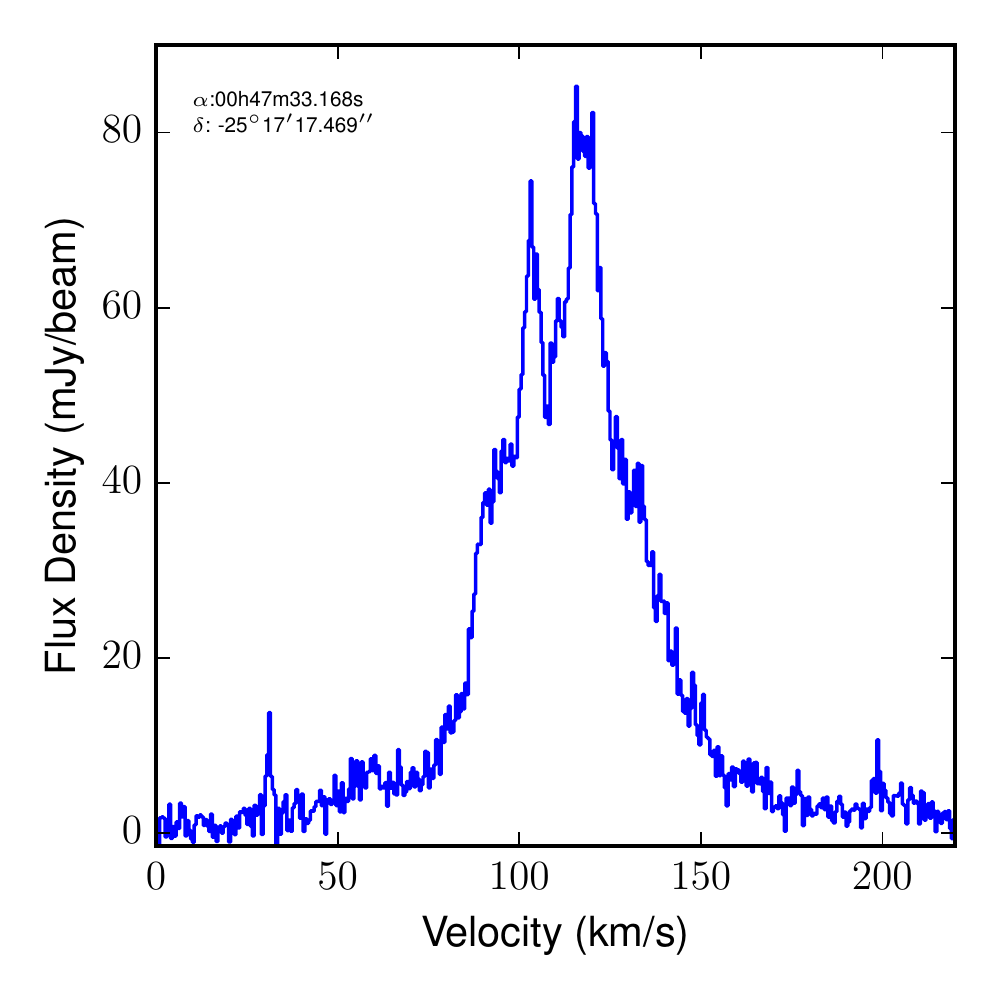}
    \caption{Spectrum of the nuclear kilomaser W1. It shows many  velocity components. Its position is shown in the upper left corner.}
    \label{fig:W1spec}
\end{figure}

We have identified 13 spatially resolved locations with \water\ masers in NGC\,253. These locations are plotted in Figure \ref{fig:loc}. The maser velocities span a range from 8$-$347\,\kms. We colour code the masers according to their velocities. Masers identified in green have velocities between 170$-$300\,\kms\ consistent with the molecular bar (e.g., \citealp{Leroy2013,Gorski2017}). Velocities $<170$\,\kms\ are plotted in blue and velocities $>300$\,\kms\ are plotted in red.  We have detected the nuclear kilomaser (e.g., \citealp{Henkel2004,Brunthaler2009,Gorski2017}; W1 in \citealt{Gorski2017}), indicated by a star-sign in Figure\,\ref{fig:loc}, and 12 other stellar class masers.  Eight of these masers lie along the molecular bar, with one spatially displaced $\sim3.2$\arcsec\ (54\,pc) to the southeast of the molecular bar, and three in other locations in the disk (Figure \ref{fig:loc}). We show the spectrum of each maser in Figure \ref{fig:waterspec} except W1, the bright kilomaser, which is shown in Figure \ref{fig:W1spec}. In the cases where the line is resolved we fit the spectrum with a Gaussian. Fluxes and flux densities are corrected for the primary beam attenuation. The maser properties are recorded in Table \ref{tab:watermasers}.
All the sources have been given a number, except W1, and we have named the sources according to the International Astronomical Union specifications \citep{Lortet1994} . 
Generally the primary beam correction is small, $< 1\%$, however for the three masers not in the nuclear starburst it can be as high as 40\%. The spectrum of the nuclear kilomaser, W1, could not be fit with a simple Gaussian or combination of Gaussians. The spectrum of W1 shows many velocity components spanning $\sim170$\,\kms. We estimated the total integrated flux of W1 by making an integrated flux map excluding 10\% of the channels at the edge of the sub-band. The total integrated flux is 3.60$\pm$0.05\,Jy\,\kms\ for a luminosity of 1.02$\pm$0.01\,\Lsun. 

\begin{table*}
\centering
\caption{ Stellar-class \water\ maser properties}
\label{tab:watermasers}
    \begin{tabular}{lllllllll}
    \hline
   \# & Name & RA (J2000)	& DEC(J2000) & $\int S d\nu$ & $V_{LSRK}$ 	& $V_{FWHM}$ 	& $S_{peak}$ & Luminosity\\
     & & 00$^h$47$^m$	& -25{\deg}~17{\arcmin} & (mJy\,\kms)	& (\kms) & (\kms) & (mJy) & $10^{-3}$\,L$_\odot$ \\
    \hline
    1&		WM\,004734.9-251657.4 			& 34.915$^s$  		& 16$^\prime$57.412\arcsec    	& \,5.4$\pm$1.6         & 211.5         		& $<$\,0.5      	& 11.1$\pm$1.6         & \,1.5$\pm$0.4\\
    2&		WM\,004733.7-251808.0 			& 33.787$^s$  		& 18$^\prime$8.031\arcsec     	& \,7.2$\pm$2.2         &  8.0          		& $<$\,0.5      	& 14.4$\pm$2.2         & \,2.1$\pm$0.6\\
    3&		WM\,004733.5-251713.7 			& 33.557$^s$  		& 13.769\arcsec    			& 12.1$\pm$1.8  	& 172.2$\pm$0.1 	& 1.6$\pm$0.3  & 7.4$\pm$1.0   	& 3.4$\pm$0.5\\
    4&		WM\,004733.2-251715.4 			& 33.296$^s$  		& 15.449\arcsec   			& 7.9$\pm$1.5   	& 174.5$\pm$0.1 	& 1.1$\pm$0.3  & 6.8$\pm$1.3   	& 2.3$\pm$0.4\\
    5&		WM\,004733.2-251720.6 			& 33.288$^s$  		& 20.699\arcsec    			& 6.4$\pm$1.1   	& 71.9$\pm$0.1  	& 0.8$\pm$0.2  & 7.9$\pm$1.5   	& 1.8$\pm$0.3\\
    6&		WM\,004733.2-251715.6 			& 33.284$^s$  		& 15.610\arcsec    			& 34.2$\pm$4.5  	& 218.5$\pm$0.7 	& 10.5$\pm$1.7  	& 3.0$\pm$0.4   	& 9.7$\pm$1.3\\
    W1& 	WM\,004733.1-251717.5			& 33.168$^s$		& 17.469\arcsec			& 3.60$\pm$0.05	& 116.0$\pm$0.1	& $-$		&85.2$\pm$1.3			&1.02$\pm$0.01\\
    7&		WM\,004733.1-251718.0 			& 33.113$^s$  		& 18.089\arcsec    			& 16.6$\pm$1.7  	& 157.0$\pm$0.1 	& 1.5$\pm$0.2  & 10.5$\pm$1.1  	& 4.7$\pm$0.5\\
    8&		WM\,004733.0-251717.6 			& 33.076$^s$  		& 17.669\arcsec    			& 25.4$\pm$2.8  	& 347.0$\pm$0.2 	& 3.9$\pm$0.5  &6.1$\pm$0.7    	& 7.2$\pm$0.8 \\
    9&		WM\,004733.0-251719.4 			& 33.001$^s$  		& 19.496\arcsec    			& 10.3$\pm$1.6  	& 207.4$\pm$0.1 	& 1.3$\pm$0.2  & 7.5$\pm$1.1   	& 2.9$\pm$0.5\\
    10&	WM\,004732.8-251721.1			& 32.836$^s$  		& 21.089\arcsec    			& 10.7$\pm$1.6  	& 310.8$\pm$0.1 	& 1.3$\pm$0.2  & 8.1$\pm$1.2   	& 3.0$\pm$0.5\\
    11&	WM\,004732.8-251721.0 			& 32.833$^s$  		& 21.039\arcsec   			& 12.3$\pm$1.7  	& 294.6$\pm$0.1 	& 1.4$\pm$0.2  & 8.4$\pm$1.2   	& 3.5$\pm$0.5\\
    12&	WM\,004732.0-251652.3 			& 32.052$^s$  		& 16$^\prime$52.355 		& \,5.2$\pm$1.5         & 265.5         		& $<$\,0.5      	& 10.4$\pm$1.5         & 1.5$\pm$0.4\\
     \hline
    \end{tabular}
    \tablenotetext {}{The typical positional uncertainty is 0.02\arcsec}
\end{table*}

\subsection{\methanol\ Masers}

\begin{figure}
    \centering
    \includegraphics[width=0.495\textwidth]{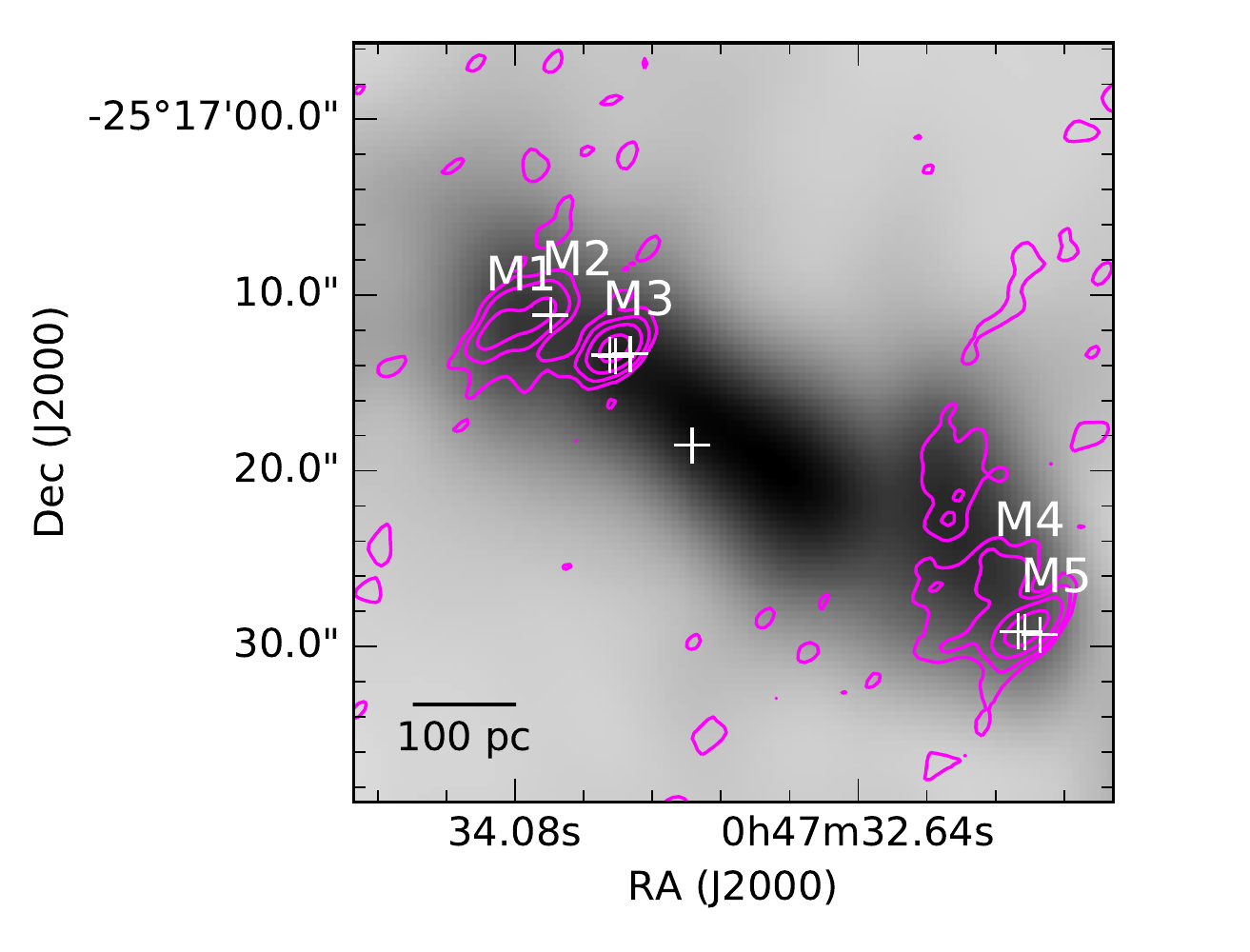}
    \caption{Positions of the eight \methanol\ masers we have detected in NGC\,253. The grayscale image shows $\sim4$\arcsec\ resolution $^{12}$CO\,($1-0$) emission from \citet{Bolatto2013}. The magenta contours show the VLA compact configuration (D-configuration) \methanol\ emission from \citet{Gorski2017}. The contour levels are 10, 20, 40 and 80 mJy\,beam$^-1$\,\kms. Masers detected in this paper are shown as plus signs.}
    \label{fig:methanol_loc}
\end{figure}

\begin{figure*}
    \centering
    \includegraphics[width=0.95\textwidth]{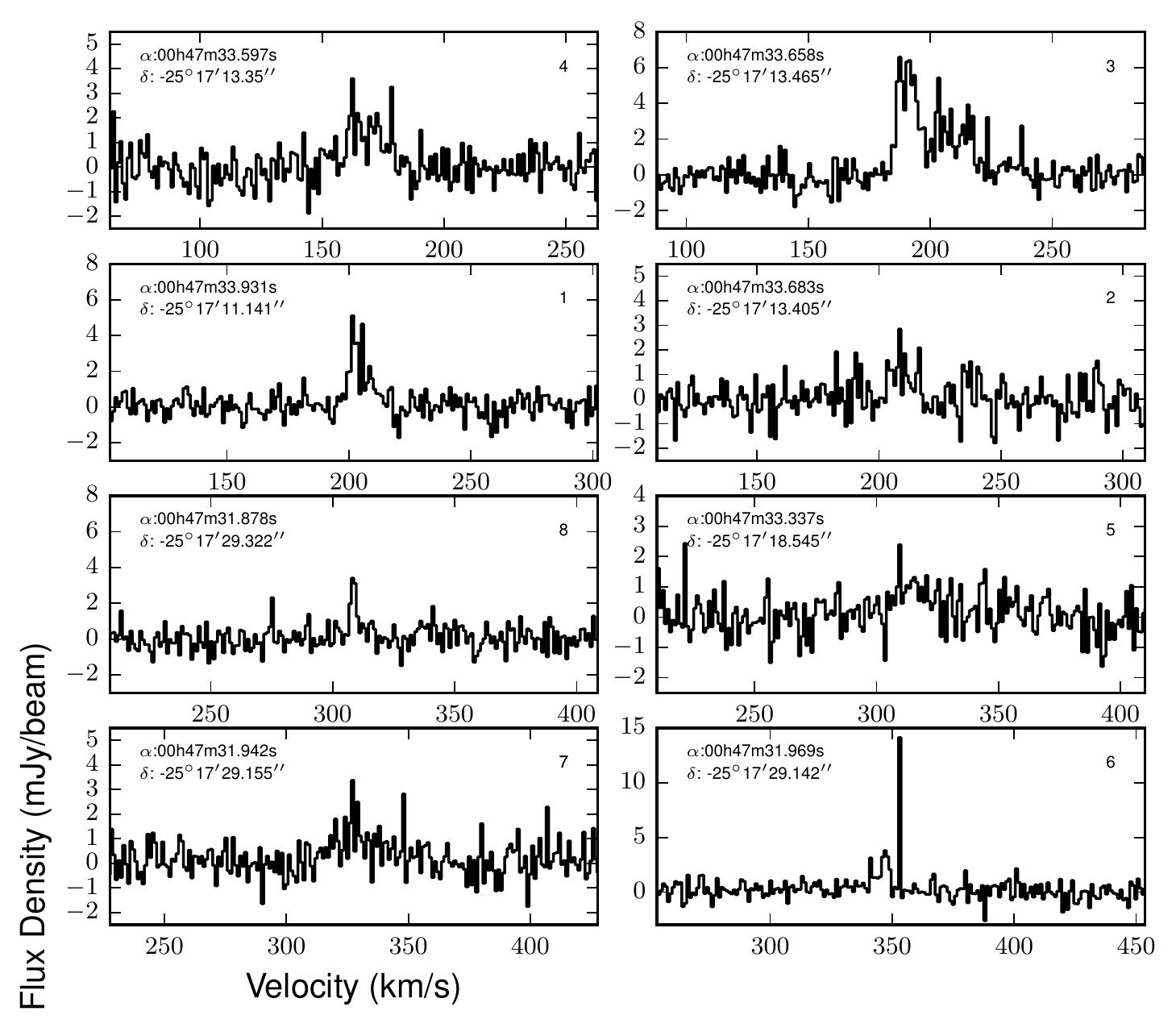}
    \caption{Spectra of the eight  class I  \methanol\ masers detected. Their coordinates are shown in the upper left corner of each panel. These spectra are not primary beam corrected. }
    \label{fig:methanolspec}
\end{figure*}

We detect seven compact masers at the locations of M2, M3, and M5 from \citet{Gorski2017} or MM6, MM4, or MM1 in \citet{Ellingsen2017} respectively, and one new maser.  Four of these sources were not detected by \citet{Chen2018}. With an rms of 20\,mJy\,beam$^{-1}$\,\kms\ (channel width of 8.3\,\kms) it is unlikely that they would have detected the fainter sources.  The locations of all the \methanol\ masers are shown in Figure \ref{fig:methanol_loc}. The extracted spectra are shown in Figure \ref{fig:methanolspec} and the properties are listed in Table \ref{tab:methanolmasers}. 

The total integrated \methanol\ flux from  \citet{Ellingsen2014} is 5.8\,Jy\,\kms\ whereas we have measured $\sim0.435$\,Jy\,\kms. The \citet{Ellingsen2014} observations have a rms noise per channel $\lesssim$0.8\,mJy\,beam$^{-1}$, a synthesized beam of 8.0\arcsec$\times$4.2\arcsec, and minimum and maximum baselines of 61 and 192\,m respectively. This means 92\% of the flux is resolved out on $\sim0.1$\arcsec\ scales if these masers are not variable. If all the emission is from compact maser sites, we would expect to recover all the flux, so this low percentage suggests most of the emission from \citet{Ellingsen2014} may be thermal and spatially extended, or there is relatively diffuse maser emission as in Galactic star forming regions like the W3(OH) region. \citealt{Menten1992} and \citealt{Harvey2006} shows maser emission over two orders of magnitude  in spatial scale, from $\sim$0.1-0.001\arcsec, in the W3(OH) region. The brightness temperatures constrain this issue somewhat. We detect a range of peak flux density values in the range 1.1$-$16.5\,mJy or brightness temperatures of approximately 60$-$1500\,K The molecular gas is estimated to have two temperature components, a cool 57$\pm$4\,K component and a warm 134$\pm$8\,K component \citep{Gorski2017}. It is possible that the less luminous sources could be thermal emission, though as these sources are unresolved, the brightness temperatures are lower limits. In general the measured properties of the \methanol\ masers are in good agreement with \citet{Chen2018}.

\begin{table*}
\centering
\caption{ \methanol\ Maser Properties}
\label{tab:methanolmasers}
    \begin{tabular}{lllllllllll}
    \hline
    \#& Name & Label$^A$ & RA (J2000)	& DEC(J2000) & $\int S d\nu$ & $V_{LSRK}$ 	& $V_{FWHM}$ 	& $S_{peak}$ & Luminosity\\
     & &   & 00$^h$47$^m$	& -25{\deg}~17{\arcmin} & (mJy\,\kms)	& (\kms) & (\kms) & (mJy) & $10^{-3}L_\odot$ \\
    \hline
    1&	 	MM\,004733.93-251711.14 		& MM6	& 33.931$^s$   & 11.141\arcsec   & 38.4$\pm$3.4  	& 204.0$\pm$0.4     & 9.1$\pm$1.0   		& 3.8$\pm$0.4   	& 17.6$\pm$1.5\\
    2&	 	MM\,004733.68-251713.40 		& -    	& 33.683$^s$   & 13.405\arcsec    & 17.1$\pm$3.3  	& 209.5$\pm$1.0     & 10.3$\pm$2.5  	& 1.6$\pm$0.3   	& 7.9$\pm$1.6\\
    3a&	MM\,004733.65-251713.45		& MM4	& 33.658$^s$   & 13.456\arcsec    & 116.5$\pm$20.7 		& 207.2$\pm$1.7     & 31.4$\pm$5.6  	& 4.9$\pm$0.4   	& 53.9$\pm$9.6\\
    3b&		 						& MM4   	& 33.658$^s$  & 13.456\arcsec    & 82.4$\pm$10.3	 	& 189.3$\pm$0.5     & 12.8$\pm$1.6	  	& 8.2$\pm$0.9   	&36.7$\pm$4.7 \\
    4&	 	MM\,004733.57-251713.35 		& -    	& 33.579$^s$   & 13.350\arcsec    & 37.2$\pm$4.7  	& 168.3$\pm$1.2     & 18.9$\pm$2.9  	& 1.7$\pm$0.2   	& 17.2$\pm$2.2\\
    5&		MM\,004733.33-251718.54 		& -          	& 33.337$^s$   & 18.545\arcsec    & 20.1$\pm$3.8  	& 315.4$\pm$1.6     & 17.0$\pm$3.9  	& 1.1$\pm$0.2   	& 9.3$\pm$1.8\\
    6a&      MM\,004731.96-251729.14 		& MM2    	& 31.969$^s$   & 29.142\arcsec    & 43.7$\pm$11.1  	& 348.4$\pm$0.8     & 10.3$\pm$2.0  	& 4.1$\pm$0.7  	& 20.2$\pm$5.7\\
    6b&								& MM2    	& 31.969$^s$   & 29.142\arcsec    & 19.4$\pm$1.8       & 353.5             		& $<$\,1.0      		& 19.4$\pm$1.7         & $<$\,8.9$\pm$0.8 \\
    7&		MM\,004731.94-251729.15 		& MM1    	& 31.942$^s$   & 29.155\arcsec    & 45.8$\pm$6.7  	& 330.3$\pm$2.0     & 26.6$\pm$4.7  	& 1.6$\pm$0.2   	& 24.9$\pm$3.1\\
    8&		MM\,004731.87-251729.32 		& -	       	& 31.878$^s$   & 29.322\arcsec    & 14.8$\pm$2.1  	& 309.0$\pm$0.2     & 2.8$\pm$0.5   		& 5.1$\pm$1.2   	& 6.8$\pm$1.0 \\
   \hline
   \end{tabular}
\tablenotetext {A}{\citep{Ellingsen2017}}
 \tablenotetext {}{The typical positional uncertainty is 0.02\arcsec}
\end{table*}

\section{Discussion}
\label{section disc}

 We detect thirteen water masers in NGC\,253: twelve stellar class masers and one kilomaser. We also detect eight class I \methanol\ 36\,GHz masers. We will interpret these results using models from \citet{Strickland2002} that describe the relationship between the nuclear starburst and the outflow. We also discuss the association of \water\ masers with compact radio sources and forming super star clusters (SSCs) described in \citet{Leroy2018}.
 
 \subsection{\water\ Masers and The Outflow}
 
 All the \water\ masers we have detected have a luminosity $<0.1$\,\Lsun\ classifying them as stellar class masers with the exception of the nuclear kilomaser W1. \citet{Palagi1993} suggest that the maximum luminosity of  evolved-star water masers is $\sim10^{-4}$\,\Lsun, hence all our detected masers are likely star formation related. \citet{Gorski2017} showed possible evidence of an \water\ maser extension along the minor axis of the galaxy. This suggested either a jet (e.g., \citealp{Peck2003}) from an active galactic nucleus (AGN) or entrained star-forming material in the outflow. In the subarcsecond resolution observations we provide in this paper we do not see masers organized along the minor axis of the galaxy in  a jet like extension suggested in \citet{Gorski2017}.  Their extension is likely an artifact resulting from generating a super-resolved image cube. This reinforces the picture that NGC\,253's nuclear environment has a pure starburst nature \citep{Brunthaler2009}.  
 
From \citet{Leroy2013} and \citet{Gorski2017} we see that the molecular gas in the central 300\,pc, which is associated with the molecular bar, has velocities in the range of $\sim$170$-$300\,\kms. In Figure \ref{fig:loc} we colour code masers with these velocities. Four of the nine masers in the nuclear 300\,pc have velocities inconsistent with the molecular bar. \citet{Westmoquette2011} show that the southeast hot ionized outflow has velocities $\sim100\pm50$\,\kms\ blueshifted with respect to the systemic velocity  of NGC\,253. The redshifted side, which is pointed away from the observer, was not detected. It is likely obscured by the foreground disk. Masers with velocities of 35$-$185\,\kms\ could be associated with the ionized outflow. As a precaution we only consider masers with velocities $<170$\,\kms\ to be part of the ionized outflow as this velocity range does not overlap with the molecular bar. W1 and three other stellar \water\ masers are in this range. The \water\ sources 2  and 5 do not lie in the molecular bar but to the southeast of the nuclear starburst. It is important to note that water masers often have velocities offset from the systemic velocity of their associated system.  In the Milky Way the largest offset is $\sim$100\,\kms\ \citep{Titmarsh2013}, but the strongest emission is within 10\,\kms\ of the systemic velocity of the system \citep{Breen2011}. 

We interpret the velocities and positions of the \water\ masers to indicate entrainment of star-forming molecular gas in the hot ionized outflow of NGC\,253. 
This supports models (b) and (c) from \citet{Strickland2002} that show entrainment of the ISM from the disk. 
The redshifted masers, \water\ 8 and 10, would represent the receding side of the outflow in this picture.   
The receding outflow is difficult to observe in the center of the galaxy because optical, ultraviolet, and X-ray observations are obscured by the disk. 
\citet{Bolatto2013} measure the velocity of the receding outflow to have velocities of $\sim$240-400\,\kms\ from $^{12}\rm{CO}(J=1\to0)$ emission, consistent with the velocity of \water\ source 8.

The nuclear kilomaser, W1, shows an increase in luminosity from 0.67\,\Lsun\ from \citet{Gorski2017} to 1.0\,\Lsun\ a \%\,42 increase. The features W1c and W3 in \citet{Gorski2017} was not detected.  The W2 feature appeared in \citet{Gorski2017} as a broad faint feature with several possible sources. Here we present many narrow spectral features within 100\,kms of W2. Though we do not see any sources related to the peaks in W2 from \citet{Gorski2017}. This is unsurprising as water masers are notoriously variable and can vary on time scales of days, weeks, or months (e.g., \citealp{Braatz1996, Claussen1996, Felli2007,Breen2013}). This shows that there are many variable water masers in the nuclear region of NGC\,253.

\subsection{Super Star Clusters and Supernovae}

\begin{figure*}
    \centering
    \includegraphics[width=0.95\textwidth]{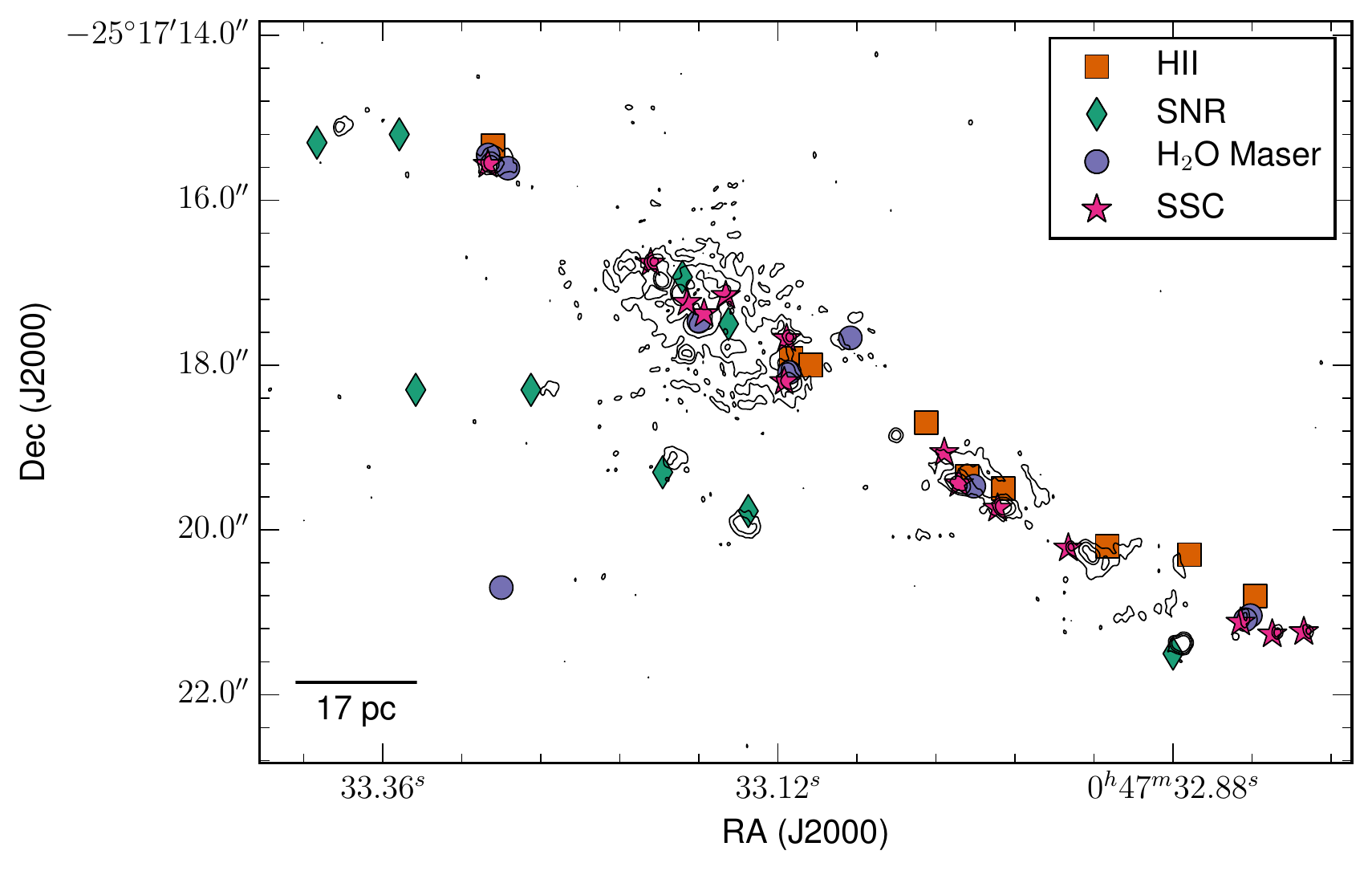}
    \caption{  The 33.5\,GHz continuum image. The contours show intervals of 3, 9, and 24 times 23\,$\mu$Jy beam$^{-1}$. \HII\ regions and SNRs from \citet{Ulvestad1997} are respectively labeled with squares and diamonds. SSCs from \citet{Leroy2018} are indicated by stars and \water\ masers by circles.}
    \label{fig:33_cont}
\end{figure*}

Five water masers are spatially coincident, within 0.23\arcsec, or one VLA synthesized beam width at 22\,GHz, with super star clusters (SSCs). \citet{Leroy2018} has identified 14 candidate super star clusters with stellar and gas masses $\gtrsim10^5$\,\Msun\ in the central 200\,pc of NGC\,253. They accomplished this by comparing 350\,GHz  with 36\,GHz  continuum emission.  They estimated dynamical masses from the widths of the CS\,($7-6$) and H$^{13}$CN($4-3$) lines, and  the stellar masses from the ionizing photon rate (\citealp{Murphy2011} and \citealp{Leitherer1999}). Of these 14 sources, numbers 3,6,9,11 and 14 from \citet{Leroy2018} have associated stellar class \water\ maser emission indicating strong star formation.  \water\ source 11 is closest to the nuclear kilomaser suggesting particularly strong star formation at this location. Figure \ref{fig:33_cont} shows the location of the SSCs relative to our water masers, and supernova remnants (SNRs) and \HII\ regions from \citet{Ulvestad1997}. 

While not all the radio continuum sources show \water\ maser emission, only sources identified as \HII\ regions in \citet{Ulvestad1997} show \water\ maser emission within one synthesized beam width.   Figure \ref{fig:33_cont} shows the 33.5\,GHz continuum image. At these frequencies the radio continuum is dominated by free-free emission that is a result of ionizing radiation from young stars, and it has a relatively flat spectral index ($\sim0.2$; e.g.,  \citealp{Condon1992,Murphy2011,Murphy2018}). Continuum sources with spectral index $\alpha<-0.4$ ($S\propto \nu^{+\alpha}$) in \citet{Ulvestad1997} show no water maser emission. This is similar to Milky Way studies (e.g., \citealp{Claussen1999, Woodall2007}) that show that the shocks in supernova remnants do not excite the \water\ maser line.  \citep{Claussen1999} suggest that unusually high densities ($>10^5$\,cm$^{-3}$) or shock velocities ($v_s>50$\,\kms) could potentially excite water masers. This is consistent with what we know of the  molecular clouds in NGC\,253. The density of the molecular clouds is $n_{\rm H_2}\sim2000$\,cm$^{-3}$ \citep{Leroy2015} or $n_{\rm H_2}\sim10^{4.5}$\,cm$^{-3}$\citep{Meier2015} in the denser regions. They are likely not dense enough for the excitation of 22\,GHz \water\ masers via interaction with SNR.

\water\ masers are variable, thus we would not necessarily expect \water\ maser detections to be associated with all SSC candidates. For example we do not detect the maser  W3 from \citet{Gorski2017}, however, we detect \water$-2$ from \citet{Henkel2004} with a flux of 12.1$\pm$1.8\,mJy\,\kms. This maser was not detected in \citet{Gorski2017} indicating an increase in brightness between 2013 and 2016, but a $\sim90\%$ decrease since it was first detected in 2002. Our results reinforce the interpretation of the emission from NGC253 as due to the starburst and not an AGN, and show that monitoring of maser sites in the nucleus could reveal more sites of intense star formation. 

\subsection{\methanol\ Masers}

\citet{Chen2018} showed that the compact 36\,GHz \methanol\ sources are mostly masers with brightness temperatures of $\sim1000$\,K. However, most of the emission is resolved out in subarcsecond-resolution interferometric observations.

In the Milky Way surveys of class I \methanol\ masers, where supernova remnants interact with molecular clouds (e.g., \citealp{Pihlstrom2014} and \citealp{McEwen2016}), are revealed to have linewidths of $\sim1$\,\kms.  Typical class I stellar \methanol\ masers have FWHMs $\sim0.5$\,\kms\ \citep{Voronkov2014}. All of our linewidths are greater than 1\,\kms\ with the exception of the bright component MM\,6a (\citealp{Ellingsen2017}; MM1) at a velocity of 353.5\,\kms. This suggests either a potentially different origin for most of our sources or we are averaging over many sources in the $\sim2$\,pc sythensized beam.  \citet{Ellingsen2017} compares the 36\,GHz \methanol\ maser with 44.1\,GHz \methanol\ maser. In both cases the luminosity is much greater than what would be expected, for high mass star formation. Due to the proximity of the \methanol\ masers to the inner Limblad resonances, \citet{Ellingsen2017} claim that the emission is generated through cloud-cloud collisions. Our results are mostly consistent with this interpretation. The exception is the source MM\,6a  which is likely stellar in origin as the line is unresolved ($V_{FWHM}<1$\kms). 

\citet{Leurini2016} estimate that class I \methanol\ masers should not vary in brightness on time scales of $\sim15$\,yrs should the maser be saturated. The maser spot has a size of $\sim100$\,AU, and a shock velocity of $\sim$30\,kms. The time between the observations presented in \citet{Chen2018} (August 2015) and this paper (October 2016) is $\sim1$\,yr. Of the four masers detected in \citet{Chen2018} we measure the same fluxes within the uncertainties. Half of our masers would be too weak to detect with their sensitivity, so it is unknown if they are variable. The time between the observations from \citet{Gorski2017} and this paper's observations is $\sim3$\,yrs, but the two sets of observations are in different array configurations meaning different spatial frequencies were sampled. This makes determining variability impossible as most emission is resolved out. However as the scale of the emission is $\gg100$\,AU, we would not expect any maser variability unless the masers are not saturated. 

There is a dearth of \methanol\ maser emission within the central 300\,pc of NGC\,253. The one maser we detect in this region is fairly weak with a peak flux density of 1.1\,mJy. It is possible that this is because the density of molecular gas rises in the galaxy center \citep{Leroy2015} and \methanol\ maser emission could be quenched at densities $>10^6$\,cm$^{-3}$(e.g., \citealp{Menten1991,McEwen2014}).  \methanol\ source 5 may be an isolated star-forming region obscured by dust, or a spurious site of lower density. It is also possible that methanol is depleted due to a stronger ultraviolet radiation field.

\citet{Ellingsen2014} suggests that the sites of the 36\,GHz methanol masers indicate the edges of the molecular outflow. This could provide evidence for model (d) of \citet{Strickland2002} showing sites of swept up molecular gas or collimation sites of the outflow. The west group of methanol masers has velocities $\sim300-350$\,\kms and the east has velocities $\sim170-210$\,\kms. These velocities are more consistent with the molecular bar and molecular outflow ( east: $40-140$\,\kms, west: $70-250$\,\kms\  \citealp{Bolatto2013}), than that of the hot-ionized outflow ($<170$\,\kms\ \citealp{Westmoquette2011} ). We do not observe any obvious indication that these are related to the molecular outflow of NGC\,253. 

\section{Summary}
\label{section sum}

We have presented subarcsecond observations of the 22\,GHz \water\ maser and 36\,GHz \methanol\ maser of the nuclear starburst in NGC\,253. We have interpreted the results through the models of outflows in \citet{Strickland2002} and compared to the formation of SSCs of \citet{Leroy2018}. We have found the following:

\begin{enumerate}
    \item We have detected thirteen water masers in NGC\,253 and provided evidence for entrainment of dense star-forming material in the hot ionized outflow of NGC\,253. The minor axis extension of \water\ maser emission from \citet{Gorski2017} is not related to an AGN. This was inferred from the lack of an extended structure and detection of only discrete sites of maser emission.
    \item We have shown that \water\ masers are positionally  associated with sites of strong star formation, and possible super star cluster formation, but not with supernova remnants. The nuclear kilomaser is associated with a forming super star cluster. 
    \item We detected eight sites with 36\,GHz \methanol\ maser emission. These sites are located at the edges of the molecular bar. They are unusually luminous compared to \methanol\ masers in star-forming regions. We are in agreement with \citet{Chen2018} that cloud-cloud collisions are the likely sources of these masers.  It is possible that these are collimation sites of the galactic wind, however the orientation of NGC\,253 makes this very difficult to determine.  We detect one weak \methanol\ maser in the centre of NGC\,253, and a narrow spectral component towards MM\,6a  suggesting a possible star-formation origin.
\end{enumerate}

The detection of star-forming dense molecular gas with velocities peculiar to the molecular bar provides evidence for the \citet {Strickland2002} models of outflows that entrain dense gas from the disk. 
The detection of water masers shows that that the entrained gas can form stars. 
From our \methanol\ maser observations we cannot rule out the model where gas is swept up from the disk generating a cool shell around the hot ionized outflow. 
It is possible that neither of these models completely describe galactic winds. \citet{Bolatto2013} show ample evidence for a cool molecular shell about the hot ionized outflow. 
Therefore combined models of entrained star forming gas and swept up cool molecular shells more accurately describe the outflow process than either model independently.

\section*{Acknowledgements}
We would like to thank Pauline Barmby for her helpful and insightful comments on the draft.

The National Radio Astronomy Observatory is a facility of the National Science Foundation operated under cooperative agreement by Associated Universities, Inc.

This research has made use of the NASA/IPAC Extragalactic Database (NED) and NASA/IPAC Infrared Science Archive, which is maintained by the Jet Propulsion Laboratory, Caltech, under contract with the National Aeronautics and Space Administration (NASA) and NASA's Astrophysical Data System Abstract Service (ADS).


\end{document}